\documentclass[12pt]{iopart}
\usepackage[dvips]{graphicx}
\usepackage{ulem}
\usepackage{color}
\newcommand{\ds}{\displaystyle}

\begin{document}

\title{Mean joint residence time of two Brownian particles in a sphere} 

\author{O. B\'{e}nichou\dag\footnote[3]{To whom correspondence should be addressed (benichou@lptl.jussieu.fr)}\ M. Coppey\dag\ J. Klafter\ddag\ M. Moreau\dag\ and G. Oshanin\dag}

\address{\dag\ Laboratoire de Physique Th{\' e}orique de la Mati{\` e}re Condens{\' e}e,
Universit{\' e} Paris 6, 4 Place Jussieu, 75252 Paris, France}

\address{\ddag\ School of Chemistry, Tel Aviv University, Tel
Aviv 69978, Israel}

\begin{abstract}
We  calculate the mean joint residence time of two Brownian particles in a sphere, for  very general initial conditions. In particular, we focus on the dependence of this residence time 
as a function of the diffusion coefficients of the two particles. Our results can be  useful for describing kinetics of  bimolecular
diffusion controlled reactions activated by catalytic sites. 
\end{abstract}

\pacs{05.40.Jc, 05.40.Fb}


\maketitle

\section{Introduction}

Elementary reactive process between two molecules $A$ and $B$ usually begins when their relative positions satisfy certain relations. In the simplest case of two molecules
 with spherical symmetry, one may consider that the reactive process starts when the relative
distance of their centres is equal to the range of their reactive
potential. This is the case for instance in the celebrated model
introduced by Smoluchowski \cite{Smolu} and improved by many authors \cite{Rice}. In a more
general model, for given values of the coordinates characterizing
$A$, the reaction can only take place if the coordinates of $B$
are located inside a reactive region $\Gamma(A)$ which, obviously,
can be fairly complicated.

 The probability that $B$ has not reacted up to time $t$ is  clearly connected to the  residence time
of $B$ inside $\Gamma(A)$ up to time $t$.
This fact seems to have been first recognized by Wilemski an Fixman \cite{Wilemski}. Indeed, in their seminal work, the inverse reaction rate constant for the $A+B\to B$ diffusion limited reactions is given by the mean residence time (RT) of the particle $B$ inside $\Gamma(A)$. We address the reader for more details to the works by  Do\"{\i} \cite{Doi}, Temkin and Yakobson \cite{Temkin}, Zhou \cite{Zhou} and Rice \cite{Rice}. The residence time inside a finite domain $S$ is defined as the total time $T(t)$ spent by the Brownian particle up to time $t$ within $S$: contrary to the first exit time of $S$,  the RT accounts for an arbitrary number of exits from and enters into $S$. The properties of the RT for an individual
Brownian particle have been extensively studied \cite{Kac,Weisslivre,Blumen1,Blumen2,Agmon,Zaloj,Kallianpur,Klafter1}.

The purpose of this paper is to develop the notion of residence time in the case of a reaction between two particles, activated by immobile catalytic sites. For such reactions, $A+B+C\to B+C$, where the specie $C$ stands for the catalytic sites, the calculation of the effective reaction rate in a Wilemski-Fixman kind approach actually involves several functionals of time which can be identified as RT of Brownian particles \cite{nousWF}. One of them is the mean time spent simultaneously by the two reactants $A$ and $B$ inside a reactive region centered on a catalytic site $C$, for which the RT is averaged over all initial positions of $A$ and $B$ inside the reactive region.  

Recently \cite{Nous}, this kind of joint residence time properties has been studied for an arbitrary number
 $N$ of Brownian particles, but   only in the two-dimensional situation and for
very particular initial conditions. Here,  we focus on the case of
two particles ($N=2$). First, it will be
shown that the mean residence time can
be calculated exactly for general initial conditions in the two dimensional case. Then, in order to obtain an explicit reaction rate for the reaction $A+B+C\to B+C$ in $3$-d, we will also study the three dimensional 
situation when particles start inside the reactive region. We will show that the mean joint residence time of $A$ and $B$ on a site $C$ can be calculated explicitly in several situation of physical interest. It is remarkable, in particular, that this problem can be solved exactly in $3$ dimensions for arbitrary initial positions of the reacting particles.

\section{The model}

We consider $2$ independent Brownian particles diffusing in a $d$
dimensional space, with $d\geq2$, and we assume that they
 can only
react on an immobile catalytic site, which will be represented by
the sphere $D(O,R)$ of center $O$ and of radius $R$. We note ${\bf
r}_j(t)$ the positions of these particles at time $t$ ($j=1,2$), ${\bf
r}_j^{(0)}$ the initial positions, and $D_j$ the corresponding
diffusion coefficients. We introduce ${\bf 1}_{\cal S}$ the
indicator function of the sphere $D(O,R)$, that is
\begin{eqnarray}
{\bf 1}_{\cal S}({\bf r})=
\cases{1&if ${\bf r}\in D(0,R)$\cr0&otherwise\cr}\nonumber\\
\end{eqnarray}

The joint occupation time of the sphere $D(O,R)$ is defined by
\begin{equation}\label{tempsavectoutes}
T(t)=\int_0^t{\rm d}t'\;\left(\prod_{j=1}^2{\bf 1}_{{\cal S}}({\bf
r}_j(t'))\right)
\end{equation}

 Thus, the  mean joint occupation time at time $t$ is
\begin{equation}\label{moment}
\mu(t)=\langle \left(T(t)\right)  \rangle
=\int_0^t\left(\prod_{j=1}^2\langle {\bf 1}_{{\cal S}}({\bf
r}_j(t'))\rangle _j\right)\;{\rm d}t',
\end{equation}
that is
\begin{equation}
\mu(t)=\int_0^t\prod_{j=1}^2 \left( \int {\rm d} {\bf r}_j {\bf
1}_{{\cal S}}({\bf r}_j) \frac{1}{(4\pi D_j t')^{d/2}}\exp\left(-\frac{||{\bf
r}_j-{\bf r}_j^{(0)}||^2}{4D_j t'}\right)\right) \;{\rm d}t'
\end{equation}

From now on, we will limit ourselves to the long time limiting value of
$\mu(t)$, which will be simply denoted as $\mu$
\begin{equation}
\label{base}
\mu\equiv\lim_{t\to\infty}\mu(t)=\frac{1}{(16\pi^2D_1D_2)^{d/2}}\int_0^\infty{\rm
d} y y^{d-2}\prod_{j=1}^2 \left( \int {\rm d} {\bf r}_j {\bf
1}_{{\cal S}}({\bf r}_j) \exp\left(-y\frac{||{\bf r}_j-{\bf
r}_j^{(0)}||^2}{4D_j }\right)\right)
\end{equation}
where we did the change of variables $y=1/t'$ in the integral.

In the Wilemski-Fixman approach of the bimolecular reactions $A+B\to B$ \cite{Wilemski,Doi,Temkin}, the quantity which is actually involved in the expression of the kinetic constant is the mean RT $\langle \mu \rangle$ of a fictive particle inside a reactive sphere of radius $r_A+r_B$, averaged over the initial position
of the particle inside the region, diffusing with $D=D_A+D_B$. 
More precisely, the overall rate constant $k$ is given
by the inverse addition law 
\begin{equation}
\frac{1}{k}=\frac{1}{k_D}+\frac{1}{k_A}
\end{equation}
where $k_D$ is the contribution due to diffusion of particles and $k_A$ is associated to the reactive process itself. It is shown that 
\begin{equation}
k_D=\frac{V}{\langle \mu \rangle},
\label{rate}
\end{equation}
 where $V$ is the reaction zone volume.

Note that this Wilemski-Fixman approach actually appears as the leading order of a perturbative theory with respect to $k_A$ which, as an unperturbed motion, takes the propagation of independent 
particles\cite{Weiss}. In the framework of this approximation, the mean residence time $\langle \mu \rangle$ is calculated by considering that the particles perform independent Brownian motions.

Our main purpose here is to compute the mean value $\langle \mu \rangle$ for the present catalytically activated reaction $A+B+C\to B+C$, which now involves two diffusing particles and a disjoint reactive region. Therefore, $\langle \mu \rangle$ will now be called the mean \textit{joint} residence time inside a region. The sequel of the
paper is structured as follows. In section \ref{origine}, we present results
concerning the case when all the particles are initially at the
origin $O$. These expressions are valid for an arbitrary dimension
of space $d$. Next, we study more general initial conditions in
dimensions $d=2$ and $d=3$ which are the physically relevant
dimensions of the problem.  In section  \ref{dimensiondeux}, we discuss the $d=2$
situation in two special cases: firstly, when all the particles
are at $t=0$ on the boundary of the sphere $D(O,R)$. Secondly,
when one of the two particles is initially at the origin while the
other one can be everywhere. In section \ref{dimensiontrois}, we solve the $d=3$
situation in full generality, that is for arbitrary initial
conditions of the diffusing particles. Therefore, this result gives the mean joint residence time involved in the bimolecular reaction activated by catalytic sites, whereas the other initial conditions mentioned previously can be related to specific reactive processes.

\section{d-dimensional case when the two particles are initially at the origin}\label{origine}

Physically, this situation corresponds, for example, to a dissociation recombination catalytically induced reaction.

We start by integrating  Eq. (\ref{base}) over $y$, which
leads, after rescaling of the spatial integration variables to
\begin{equation}
\mu=\frac{\sigma_d^2 R^2 \Gamma(d-1)}{\pi^d} \;I(D_1,D_2)
\end {equation}
where
\begin{equation}
I(D_1,D_2)=\int_0^{1/\sqrt{4D_1}} {\rm d}x\int_0^{1/\sqrt{4D_2}} {\rm d}y \frac{x^{d-1}y^{d-1}}{(x^2+y^2)^{d-1}}
\end{equation}
and $\sigma_d$ is the area of the d-dimensional hypersphere of unit radius.

We now proceed to the calculation of $I(D_1,D_2)$. First, we write $I(D_1,D_2)=J(D_1,D_2)+J(D_2,D_1)$ with
\begin{equation}
J(D_1,D_2)=\int_{{\cal D}}{\rm d}x {\rm d}y  \frac{x^{d-1}y^{d-1}}{(x^2+y^2)^{d-1}}
\end {equation}
and ${\cal D}$ is the domain
${\cal D}=\{x\in[0,\frac{1}{\sqrt{4D_1}}],\;y\in[0,\frac{1}{\sqrt{4D_2}}],
\;y\leq\sqrt{\frac{D_1}{D_2}}x\}$.

Changing to the new variables $z=x,\;t=y/x$ in $J(D_1,D_2)$, we obtain
\begin{eqnarray}
J(D_1,D_2)&=&\int_0^{1/\sqrt{4D_1}}{\rm d}z z\;\int_0^{\sqrt{D_1/D_2}}{\rm d} t
\frac{t^{d-1}}{(1+t^2)^{d-1}}\nonumber\\
&=&\frac{1}{8dD_1}\left(\frac{D_1}{D_2}\right)^{d/2}\;_2F_1\left(d-1,\frac{d}{2};1+\frac{d}{2};-\frac{D_1}{D_2}\right),
\end{eqnarray}
$_2F_1$ being the standard Gauss hypergeometric function.

Thus,
\begin{equation}
\mu=\frac{\sigma_d^2R^2\Gamma(d-1)}{8 d \pi^d}\left\{\frac{1}{D_1}g\left(\frac{D_1}{D_2}\right)+
\frac{1}{D_2}g\left(\frac{D_2}{D_1}\right)\right\}
\end{equation}
where
\begin{equation}
g(x)=x^{d/2}\; _2F_1\left(d-1,\frac{d}{2};1+\frac{d}{2};-x\right)
\end{equation}
Using the expression $\sigma_d=d\pi^{d/2}/\Gamma(1+d/2)$ as well as the duplication formula
\begin{equation}
\Gamma(d-1)=\frac{2^{d-2}}{\sqrt{\pi}}\Gamma\left(\frac{d-1}{2}\right)\Gamma\left(\frac{d}{2}\right)
\end{equation}
we finally find
\begin{equation}
\mu=\frac{R^2}{\sqrt{\pi}}2^{d-4}\frac{\Gamma\left(\frac{d-1}{2}\right)}{\Gamma\left(\frac{d}{2}+1\right)}
\left\{\frac{1}{D_1}g\left(\frac{D_1}{D_2}\right)+
\frac{1}{D_2}g\left(\frac{D_2}{D_1}\right)\right\}
\end{equation}

Specializing to the dimensions $d=2$ and $d=3$, we get the
following explicit expression of the mean residence time $\mu$ as
a function of $R$, $D_1$ and $D_2$ :
\begin{itemize}
\item for $d=2$ : $g(x)=\ln(1+x)$, thus
\begin{equation}
\label{d2}
\mu=\frac{R^2}{4}\left\{\frac{1}{D_1}\ln\left(1+\frac{D_1}{D_2}\right)+
\frac{1}{D_2}\ln\left(1+\frac{D_2}{D_1}\right)\right\}
\end{equation}
This result has already been obtained in ref \cite{Nous} by a slightly different approach.

\item for $d=3$ : $\displaystyle g(x)=\frac{3}{2}\left[\arctan(\sqrt{x})-\frac{\sqrt{x}}{1+x}\right]$, and
\end{itemize}
\begin{eqnarray}
\label{d3}
\displaystyle
\mu=\frac{R^2}{\pi}\left\{\frac{1}{D_1}\arctan\left(\sqrt{\frac{D_1}{D_2}}\right)
+\frac{1}{D_2}\arctan\left(\sqrt{\frac{D_2}{D_1}}\right)-\frac{1}{\sqrt{D_1D_2}}
\right\}\
\end{eqnarray}

We now  briefly comment on the expressions (\ref{d2}) and
(\ref{d3}) in the special case when the particle 1 moves much
faster than the particle 2, that is when $D_1\gg D_2$.

In that case, the two dimensional expression Eq.(\ref{d2}) becomes
\begin{equation}\label{limiteD1D2}
\mu\sim_{\frac{D_2}{D_1}\to
0}\frac{R^2}{4D_1}\ln\left(\frac{D_1}{D_2}\right).
\end{equation}
The form of this last result appears, from the physical point of
view, to be  very reasonable. As a first approximation, we can say
that this joint occupation time is given by the joint occupation
time before the first exit time of the slow particle out of the
disc. This time is of order $\displaystyle t_2=\frac{R^2}{D_2}$.
Meanwhile, the fast particle leaves the disc and comes back
several times. The time spent at each return of the particle 1
inside the disc is of order $\displaystyle t_1=\frac{R^2}{D_1}$,
and the order of magnitude of the number of returns at time $t_2$
of the particle one inside the disc is given by $\displaystyle
\ln\left(\frac{t_2}{t_1}\right)$, in the limit of long $t_2$
times. As a consequence, we expect
\begin{equation}
\mu\sim_{\frac{D_2}{D_1}\to 0}
t_1\ln\left(\frac{t_2}{t_1}\right)\sim
\frac{R^2}{D_1}\ln\left(\frac{D_1}{D_2}\right)
\end{equation}
which reproduces the dependence of $\mu$ as a function of the
diffusion coefficients $D_1$ and $D_2$ given by Eq. (\ref{d2}).

Concerning the three dimensional expression Eq.(\ref{d3}), the
limit $D_1\gg D_2$ leads to
\begin{equation}
\mu\sim_{\frac{D_2}{D_1}\to 0}\frac{R^2}{2D_1}.
\end{equation}
This is clearly the expected result: knowing that the three
dimensional Brownian motion is transient and that the particle 2
leaves the sphere after a very long time, the mean joint residence
time identifies itself to the residence time of the only particle
1, which is precisely  given by $\frac{R^2}{2D_1}$.

We now examine more general initial conditions, and we restrict
ourselves to the physically important cases corresponding to the
two-dimensional and three-dimensional situations.

\section{Two-dimensional situation}\label{dimensiondeux}

Integrating first with respect to the angular variables in Eq.
(\ref{base})
\begin{equation}\label{representationutile}
\mu=\frac{1}{4D_1D_2}\int_0^\infty {\rm d} y\prod_{j=1}^2 \left(
\int_0^R {\rm d}
 r_j r_j {\rm I}_0\left(\frac{yr_j^{(0)}r_j}{2D_j}\right)\exp\left(-\frac{1}{4D_j}y(r_j^2+(r_j^{(0)})^2)\right)\right),
\end{equation}
where ${\rm I}_0$ stands for a modified Bessel function of zeroth
order. We have not been able to compute explicitly this integral
for arbitrary initial conditions, but only in special cases: the
first one is the case when the two particles are initially on the
boundary of the disc and the second one the case when one of the
two particles is initially at the origin.

\subsection{Case when the two particles are initially on the boundary of the disc}

In that special case, the spatial integrations
can be done explicitly.
\begin{eqnarray}
&&\int_0^R {\rm d}
 r_j r_j {\rm I}_0\left(\frac{yRr_j}{2D_j}\right)
 \exp\left(-\frac{1}{4D_j}y(r_j^2+R^2)\right)\nonumber\\
 &
=&R^2\;\exp\left(-\frac{yR^2}{4D_j}\right)\int_0^1{\rm d}u_j u_j {\rm
I}_0\left(\frac{yR^2u_j}{2D_j}\right)\exp\left(-\frac{yu_j^2}{4D_j}\right)\nonumber\\
&=&\frac{D_j}{y}\left(1-\exp\left(-\frac{R^2y}{2D_j}\right){\rm
I}_0\left(\frac{yR^2}{2D_j}\right)\right)
\end{eqnarray}
where we have used the fact that \cite{Grad}
\begin{equation}
\int_0^1{\rm d} x  x e^{-\alpha x^2} {\rm I}_0(2\alpha
x)=\frac{1}{4\alpha}\left(e^\alpha-e^{-\alpha}{\rm I}_0(2\alpha)\right)
\end{equation}
Finally,
\begin{equation}
\mu=\frac{1}{4}\int_0^\infty \frac{{\rm d}
y}{y^2}\prod_{j=1}^2\left(1-\exp\left(-\frac{R^2y}{2D_j}\right){\rm
I}_0\left(\frac{yR^2}{2D_j}\right)\right)
\end{equation}

In the special case when the two particles have the same diffusion
coefficient $D_1=D_2=D$, this equation becomes
\begin{equation}
\mu=\frac{R^2}{8D}\int_0^\infty\frac{{\rm d}t}{t^2}\left(1-e^{-t}{\rm
I}_0(t)\right)^2,
\end{equation}
where the last integral can be numerically evaluated as
\begin{equation}
\int_0^\infty\frac{{\rm d}t}{t^2}\left(1-e^{-t}{\rm
I}_0(t)\right)^2\approx 1.06
\end{equation}

\subsection{Case when one of the two particles is initially at the origin}

We consider now a more general situation corresponding to the case
when one the two particles is initially at the origin while the
other one can be everywhere. This situation can be a simple model for a third order reaction
between an immobile site  $C$ and particles 1 and 2, if $C$
and particle 1 initially constitute a complex molecule which
should be dissociated to react with a chemical species 2 (provided that the
 distances  from 1 and 2 to $C$ are less than $R$). 

The residence time of 1 and 2 within the catalytic region is then
\begin{eqnarray}
\mu&=&\frac{1}{4D_1D_2}\int_0^\infty {\rm d} y \left( \int_0^R {\rm d}
 r_1 r_1 \exp\left(-\frac{1}{4D_1}yr_1^2\right)\right)×\nonumber\\
&×&
 \left(\int_0^R {\rm d}
 r_2 r_2 {\rm I}_0\left(\frac{yr_2^{(0)}r_2}{2D_2}\right)
 \exp\left(-\frac{1}{4D_2}y(r_2^2+(r_2^{(0)})^2)\right)\right),
\end{eqnarray}

We perform first the integral with respect to $y$:
\begin{eqnarray}
\resizebox{\linewidth}{!}{\mbox{$\ds \int_0^\infty{\rm d}y\exp\left[-y\left(\frac{r_1^2}{4D_1}+\frac{r_2^2+(r_2^{(0)})^2}{4D_2}\right)\right]{\rm
I}_0\left(\frac{yr_2^{(0)}r_2}{2D_2}\right)=
\left[\left(\frac{r_1^2}{4D_1}+\frac{r_2^2+(r_2^{(0)})^2}{4D_2}\right)^2-\frac{(r_2r_2^{(0)})^2}{4D_2^2}\right]^{-1/2}$}}\nonumber\\
\end{eqnarray}
Performing next the integration over $r_2$ and lastly over $r_1$,
introducing the dimensionless variables $\displaystyle
x=\frac{r_2^{(0)}}{R}$ and $w=\frac{D_2}{D_1}$, we find after
straightforward but lengthy calculations that
\begin{itemize}
  \item if $x<1$
\end{itemize}
\begin{equation}\label{Eq1}
\mu=\frac{R^2}{8}\left\{\frac{1}{D_1}(1-\ln A_2)
+\frac{1}{D_2}(1-x^2-\ln A_1)
-\frac{1}{D_2}\left((1+x^2+w)^2-4x^2\right)^{1/2}\right\}
\end{equation}
\begin{itemize}
\item if $x>1$
\end{itemize}
  \begin{equation}\label{Eq2}
\mu=\frac{R^2}{8}\left\{\frac{1}{D_1}(1-\ln(A_2)
+\frac{1}{D_2}(x^2-1-\ln(x^4  A_1))
-\frac{1}{D_2}\left((1+x^2+w)^2-4x^2\right)^{1/2}\right\}
\end{equation}

where
\begin{equation}\label{Eq3}
A_1=\frac{1}{2x^4}\left\{(1+x^2+w)^2-2x^2-(1+x^2+w)\sqrt{(1+x^2+w)^2-4x^2}\right\}
\end{equation}
and
\begin{equation}\label{Eq4}
A_2=\frac{1}{2x^4}\left\{(1-x^2+w)^2+2wx^2-(1-x^2+w)\sqrt{(1+x^2+w)^2-4x^2}\right\}
\end{equation}

In the special case when the two particles have the same diffusion
coefficient $D_1=D_2=D$, this last expressions simplify in
\begin{equation}\label{Eq5}
\mu=\frac{R^2}{8D}\;f(x),
\end{equation}
where
\begin{eqnarray}\label{Eq6}
f(x)=
\cases{x^2-\sqrt{x^4+4}+\ln\left(\frac{x^4+8+4\sqrt{x^4+4}}{x^4}\right)&$1<x$\cr
2-x^2 -\sqrt{x^4+4}+\ln\left(x^4+8+4\sqrt{x^4+4}\right)&$x<1$\cr}\nonumber\\
\end{eqnarray}

In the spirit of formula (\ref{rate}), it is interesting to obtain the mean value of $\mu$ with respect
to the initial position of $r_2$. Assuming that the particle 2 is
initially uniformly distributed in the disc $D(0,R)$, we get
\begin{eqnarray}
\langle \mu \rangle &=&\frac{\int_0^R r_2^{(0)}{\rm d}r_2^{(0)}
 \int_0^{2\pi}{\rm d} \theta_2^{(0)} \;\mu({\bf r}_2^{(0)})}{\pi R^2}\nonumber\\
&=&\frac{R^2}{16}\left\{\frac{1}{D_1}b_1\left(\frac{D_2}{D_1}\right)+\frac{1}{D_2}b_2\left(\frac{D_2}{D_1}\right)\right\}
\end{eqnarray}
where
\begin{eqnarray}
b_1(x)&=&a(x)-x+2\ln\left(\frac{2(x+a(x))}{x(x+2-a(x))(-x+a(x))}\right)\nonumber\\
b_2(x)&=&-2a(x)+2\ln\left(\frac{2}{x^2+4x+2-(x+2)a(x)}\right)\nonumber\\
a(x)&=&\sqrt{(x(4+x))}
\end{eqnarray}

  In the case of equal
 diffusion coefficients, it is given by the very simple formula
\begin{equation}
\label{simple}
\langle\mu\rangle\approx 0.28 \;\frac{R^2}{D}.
\end{equation}

In the limit $D_1\gg D_2$, we obtain $\displaystyle\langle \mu \rangle \sim \frac{R^2}{8 D_1} \ln\left(\frac{D_1}{D_2}\right)$, this result admitting the same interpretation than the one given after Eq.(\ref{d2}).

\section{Three dimensional situation}\label{dimensiontrois}

We now turn to the most challenging situation, that is the three
dimensional case for arbitrary initial conditions.

Integrating first over $y$ in Eq. (\ref{base}) and denoting for
simplicity $r_1^{(0)}=v$ and $r_2^{(0)}=u$, we have
\begin{eqnarray}
\label{derniere}
\mu=\frac{1}{(16\pi^2D_1D_2)^{3/2}}\int\left(\prod_{i=1}^2 {\rm d} {\bf r}_i
{\bf 1}_{{\cal S}}({\bf r}_i)\frac{1}{\sum_{j=1}^2 \frac{||{\bf
r}_j-{\bf r}_j^{(0)}||^2}{4D_j }}\right)\nonumber\\
\resizebox{\linewidth}{!}{\mbox{$\ds=\frac{(2\pi)^2}{(16\pi D_1D_2)^{3/2}}\int_0^R{\rm d} r_1 \int_0^R{\rm d}r_2
\int_0^\pi {\rm d}\theta_1 \int_0^\pi {\rm d}\theta_2 \frac{r_1^2r_2^2\sin{\theta_1}\sin{\theta_2}}
{\left(\frac{1}{D_1}(r_1^2+v^2-2r_1v\cos{\theta_1})+\frac{1}{D_2}(r_2^2+u^2-2r_2u\cos{\theta_2})\right)^2}$}}\nonumber\\
\end{eqnarray}

It turns out that all the integrations can be performed exactly,
and $\mu$ can be expressed in terms of elementary functions of the
initial distances $v$ and $u$ of particles 1 and 2 to the center
of the catalytic site. These (rather lengthy) expressions are
listed in the Appendix.

Averaging over the initial positions of particles (which are supposed to be uniformly distributed in the reactive zone), we find
\begin{eqnarray} \label{JTR}
\langle \mu \rangle &=&
\frac{(4\pi)^2\;\int_0^R{\rm d}v v^2\;\int_0^R{\rm d}u u^2\mu(v,u)}{(4\pi R^3/3)^2}\nonumber\\
&=& \frac{R^2}{20\pi}\left\{\frac{1}{D_1}m\left(\frac{D_1}{D_2}\right)+
\frac{1}{D_2}m\left(\frac{D_2}{D_1}\right)\right\}
\end{eqnarray}
where
\begin{eqnarray}
m(x)=\frac{2-10\ln(1+x)}{x^{1/2}}-2\frac{\ln(1+x)}{x^{3/2}}+16\arctan(\sqrt{x})-\frac{7}{2} x^{1/2}
\end{eqnarray}

Note that in the special case when one of the two particles, say 2,  is immobile ($D_2=0$), 
following the Wilemski-Fixman \cite{Wilemski,Doi,Temkin} approach and  Eq.(\ref{rate}),  we obtain the expression 
\begin{equation}
k_D=\frac{5}{6} (4\pi D_1 R)
\end{equation}
given in the paper of Do\"{\i} \cite{Doi}, which differs  from the Smoluchowski constant only by the factor $5/6$.

Note also that in the special case when the particle 2 is immobile and initially within the catalytic
region, we recover the expressions given in \cite{Zaloj} for the mean residence time of a single Brownian particle within a sphere.

\section{Conclusion}

We have computed the mean joint residence time of two particles in a sphere, for  very general 
initial 
conditions. In particular, we have obtained in all cases the dependence of this residence time 
as a function of the diffusion coefficients of the two particles. 
Such results can be especially useful for describing kinetics of a bimolecular reaction activated by catalytic sites. For this reaction in three dimensions, the effective reaction rate actually involves the mean joint residence time of two reactive particles inside a region centered on a catalytic site. This mean time, being averaged over all initial positions of the two reactive particles inside the reactive region, is explicitly given by Eq.(\ref{JTR}). This is our main result, which appears as a generalization of  the expression of the reaction constant given by   Wilemski and Fixman   for a bimolecular reaction.

\section{Appendix : Expressions of $\mu$ in three dimensions }

Considering the expression (\ref{derniere} of $\mu$, we integrate over angular variables and obtain :
\begin{eqnarray}
\int_0^\pi {\rm d}\theta_1 \int_0^\pi {\rm d}\theta_2 \frac{r_1^2r_2^2\sin{\theta_1}\sin{\theta_2}}
{\left(\frac{1}{D_1}(r_1^2+v^2-2r_1v\cos{\theta_1})+\frac{1}{D_2}(r_2^2+u^2-2r_2u\cos{\theta_2})\right)^2}=\frac{D_1D_2r_1r_2}{4vu}×\nonumber\\
\resizebox{\linewidth}{!}{\mbox{$\ds ××\ln\left(\frac{(D_2(r_1^2+v^2)+D_1(r_2^2+u^2)+2D_2vr_1-2D_1ur_2)
(D_2(r_1^2+v^2)+D_1(r_2^2+u^2)-2D_2vr_1+2D_1ur_2)}
{(D_2(r_1^2+v^2)+D_1(r_2^2+u^2)-2D_2vr_1-2D_1ur_2)
(D_2(r_1^2+v^2)+D_1(r_2^2+u^2)+2D_2vr_1+2D_1ur_2)}\right)$}}\nonumber\\
\end{eqnarray}
Integrating next by parts over the radial variables $r_1$ and $r_2$, we finally find for initial positions $v$ and $u$
\begin{eqnarray}
\mu(v,u)=-\frac{1}{96\pi uv}\left\{16\frac{uvR^2}{\sqrt{D_1D_2}}+16R^3\left(\frac{u}{D_1}f(u,v;D_1,D_2)+
\frac{v}{D_2}f(v,u;D_2,D_1)\right)+\right.\nonumber\\
\resizebox{\linewidth}{!}{\mbox{$\ds+\left.8uv\left(\frac{v^2}{D_1}h(u,v;D_1,D_2)+\frac{u^2}{D_2}h(v,u;D_2,D_1)\right)
+24uvR^2\left(\frac{1}{D_2}h(u,v;D_1,D_2)+\frac{1}{D_1}h(v,u;D_2,D_1)\right)+\right.$}}\nonumber\\
\resizebox{\linewidth}{!}{\mbox{$\ds+\left.\left[D_1^2u^4+D_2^2v^4-6D_1D_2u^2v^2-6R^2(D_1^2u^2+D^2v^2-D_1D_2(u^2+v^2))-3R^4(D_1+D_2)^2
\right]\frac{\psi(u,v;D_1,D_2)}{(D_1D_2)^{3/2}}+\right.$}}\nonumber\\
+\left.\frac{8R^3}{(D_1D_2)^{3/2}}\left[D_1^2u\theta(u,v;D_1,D_2)+D_2^2v\theta(v,u;D_2,D_1)\right]
\right\}+\eta(u,v;D_1,D_2)
\end{eqnarray}
where

\begin{eqnarray}
f(u,v;D_1,D_2)&=&\arctan \left( \sqrt {{\frac {{D_2}}{{D_1}}}} \frac{\left( R+v \right)}  {\left( R-u \right) } \right) +
\arctan \left( \sqrt {{\frac {{D_2}}{{D_1}}}} \frac{\left( R+v \right)}  {\left( R+u \right) } \right)-\nonumber\\
&-&
\arctan \left( \sqrt {{\frac {{D_2}}{{D_1}}}} \frac{\left( R-v \right)}  {\left( R-u \right) } \right)-
\arctan \left( \sqrt {{\frac {{D_2}}{{D_1}}}} \frac{\left( R-v \right)}  {\left( R+u \right) } \right)
\end{eqnarray}
\begin{eqnarray}
h(u,v;D_1,D_2)&=&\arctan \left( \sqrt {{\frac {{D_1}}{{D_2}}}} \frac{\left( R+u \right)}  {\left( R-v \right) } \right) +
\arctan \left( \sqrt {{\frac {{D_1}}{{D_2}}}} \frac{\left( R+u \right)}  {\left( R+v \right) } \right)+\nonumber\\
&+&
\arctan \left( \sqrt {{\frac {{D_1}}{{D_2}}}} \frac{\left( R-u \right)}  {\left( R-v \right) } \right)+
\arctan \left( \sqrt {{\frac {{D_1}}{{D_2}}}} \frac{\left( R-u \right)}  {\left( R+v \right) } \right)
\end{eqnarray}
\begin{eqnarray}
&&\psi(u,v;D_1,D_2)=\nonumber\\
&&\resizebox{\linewidth}{!}{\mbox{$\ds\ln  \left(
\frac{((D_1+D_2)R^2+D_1u^2+D_2v^2-2D_1uR+2D_2vR)((D_1+D_2)R^2+D_1u^2+D_2v^2+2D_1uR-2D_2vR)}
{((D_1+D_2)R^2+D_1u^2+D_2v^2-2D_1uR-2D_2vR)((D_1+D_2)R^2+D_1u^2+D_2v^2+2D_1uR+2D_2vR)}
\right)$}}\nonumber\\
\end{eqnarray}
\begin{eqnarray}
&&\theta(u,v;D_1,D_2)=\nonumber\\
&&\resizebox{\linewidth}{!}{\mbox{$\ds\ln  \left(
\frac{((D_1+D_2)R^2+D_1u^2+D_2v^2-2D_1uR+2D_2vR)((D_1+D_2)R^2+D_1u^2+D_2v^2+2D_1uR+2D_2vR)}
{((D_1+D_2)R^2+D_1u^2+D_2v^2+2D_1uR-2D_2vR)((D_1+D_2)R^2+D_1u^2+D_2v^2-2D_1uR-2D_2vR)}
\right)$}}\nonumber\\
\end{eqnarray}
\begin{eqnarray}
\eta(u,v;D_1,D_2)=\cases{1/2\,{R}^{2} \left(
\frac{1}{D_1}+\frac{1}{D_2} \right) &$u<R$\  and \ $v<R$\cr
1/3\,{\frac {{R}^{3}}{u{\it D_2}}}&$R<u$\  and \ $v<R$\cr
1/3\,{\frac {{R}^{3}}{v{\it D_1}}}&$u<R$\  and \ $R<v$\cr
0&otherwise\cr}
\end{eqnarray}

\newpage
\bibliography{biblio}

\end{document}